\documentclass[sigconf]{acmart}

\usepackage{balance}

\usepackage{booktabs} 

\setcopyright{rightsretained}

\begin{document}
\title{Barriers to Active Learning for Computer Science Faculty}

\author{Jesse Eickholt}
\affiliation{%
  \institution{Central Michigan University}
  \city{Mt. Pleasant}
  \state{MI}
  \postcode{48859}
}
\email{eickh1jl@cmich.edu}

\begin{abstract}
Active learning is a proven pedagogical style that has demonstrated value by improving students'
performance and classroom experience. In spite of the evidence, adoption of active learning in
computer science remains relatively low. To identify what barriers to adoption exist, an
electronic survey was sent to 369 computer science faculty in a state in the Upper Midwest and to 78
administrators and support staff. Analysis of the responses revealed that time remained the most
commonly reported barrier for faculty that desire to change their teaching style, with 42.8\% of
faculty respondents disagreeing with the statement that they have the time they need to change
their teaching style. Administrators and support staff also indicated that time was a concern but
that otherwise faculty were aware of active learning and had the resources they need. Reported
use of active learning pedagogy was much higher among faculty that received pedagogical
training during their undergraduate or graduate studies. Given the time constraints of faculty, it
is recommended that new avenues be explored to provide future faculty with exposure to active
learning pedagogy in their undergraduate and graduate training.
\end{abstract}

%
%

\ccsdesc[500]{Social and professional topics~Computing education}

\keywords{active learning, changing teaching style, computer science faculty}

\maketitle

\section{Introduction}
Evidence-based teaching practices are practices that are known to have pedagogical value and are effective for promoting student success. If this is indeed the case, natural questions to ask are what are the adoption rates of these practices and what barriers to adoption, perceived or actual, exist? These questions are particularly pertinent for disciplines that lack diversity or capacity. By definition, evidence-based teaching practices such as active learning should address broadening participation and increasing retention.

One development in recent years is the blossoming of active learning pedagogy and
associated active learning classrooms in higher education and science, technology, engineering and mathematics (STEM) classes. A meta-analysis of the literature conducted by Freeman et al. revealed that active learning increases student performance in STEM areas and that students in active learning sections of a course were 1.5 times less likely to fail than their counterparts who took the same course in a traditional lecture format\cite{freeman2014active}. The study also showed that the students in active learning sections experienced a 6\% boost on exam scores on average and reported that the benefits of active learning cross gender boundaries. 

In spite of the reported benefits of active learning and its potential to increase the capacity of the STEM pipeline, adoption of active learning pedagogy has been limited. In a survey of electrical and computer engineering faculty conducted in 2011, reported use of active learning pedagogy ranged between 20 to 30\% depending on the technique \cite{froyd2013estimates}. In a more recent survey of computer science faculty, only 20\% of faculty reported to make use of student centered pedagogy with 38\% regularly relying on lectures for content delivery \cite{grissom2017student}. Somewhat more promising was a recent report on the climate of evidence-based teaching among STEM faculty at a large, research institution. It stated that STEM faculty reported their teaching to be more student centered \cite{landrum2017assessing}.  Thus, there seems to be growing interest in active learning and evidence-based practices but adoption could still be characterized as low.

Limited faculty adoption could be due to a number of factors. One factor is the cost of active learning classrooms and associated technology, which can easily reach hundreds of thousands of dollars \cite{park2014transformation}. Recent efforts by the community have created economical options to support active learning and active learning classrooms \cite{cole2017creating,dufresne1996classtalk}. These developments, along with the fact that active learning pedagogy can be employed to an extent regardless of the environment, have overcome one obstacle to wider adoption of active learning pedagogy. A major remaining obstacle is faculty and administration and reticence on their part to bring active learning into their classrooms. It is natural to ask why adoption rates for active learning are at their current level given their reported value in the literature. Asking and understanding the response to this question could potentially provide more computer science students with an improved classroom experience and increase retention. Increased classroom performance and an increase classroom experience could ameliorate the challenges caused by shortages of software developers and information technology specialists.

Presented here is a pilot study to investigate reported teaching style, characteristics of faculty and their view of support services.  The aim is to begin to investigate additional barriers to active learning adoption by computer science faculty and identify potential opportunities to increase adoption of this proven pedagogy.  If particular prior behaviors or attitudes correlate to resistance in the adoption of active learning, then these could constitute additional barriers that need to be addressed.  Alternatively, if there are behaviors or attitudes that correlate to increased use, then these may be a means to increase adoption.  Support services to aid faculty with changes to teaching style may also inhibit active learning or the support services offered may not fit the preferred modalities of faculty.  Differing views on the amount of support between faculty and administrators and support staff could constitute yet another barrier.  These lines of inquiry were addressed by surveying both computer science faculty and administrators and support staff.         

\section{Background}

\subsection{Active Learning}
Active learning is a pedagogical approach that puts students in control of the learning process. In particular, it focuses on activities that involve meta-cognition \cite{learn2000brain}. These activities help students master concepts by taking control of their own learning and monitoring their own progress towards meeting specific learning objectives. A student's ability to monitor his or her own progress towards mastery is key since a broad view of the literature has revealed that students engage in the classroom with a preconceived view of the world and how they believe it works, much of which may be incorrect \cite{learn2000brain}. If a student does not have time to discuss, think about and break down existing notions, then new knowledge will not be retained. Furthermore, for students to achieve competency in an area they must not only know the facts but also be able to apply them. This comes from assembling new knowledge into a workable framework \cite{learn2000brain}. There is value then in pedagogical approaches that force the student to not just view facts and attempt to retain them but to engage and challenge preconceived notations, investigate alternative explanations and make decisions on how and what to learn. These are activities that active learning promotes and a recent meta-analysis of the literature found that it is a particularly useful pedagogical approach
for STEM disciplines \cite{freeman2014active}.

Active learning has broadly been defined as a teaching practice that engages students in the learning process \cite{prince2004does}. In terms of pedagogies, there are many ways in which this type of engagement can be achieved. Some common examples are process oriented, guided instructional learning (POGIL), problem based learning, and peer instruction \cite{baepler2016guide}. Peer instruction has students fill both the role of student and teacher \cite{eberlein2008pedagogies}. This can take many forms or be performed at many levels. Students who have had previous experience could serve leaders of small workshops or one could image a student teaching a topic of interest. Problem based learning has groups working through well-defined problems \cite{barrows1996problem}. Students may have assigned roles such as scribe or investigator. They may also have to look for resources as they work towards a solution.

Aside from specific active learning styles or pedagogies, there are a number of techniques that can be used to bring active learning into the classroom regardless of the teaching style. One such technique is think-pair-share in which a question is posed to students to answer individually \cite{kagan1989structural}. After pondering a response, a student must pair with a neighbor and discuss to reach a consensus. Another technique is the minute paper \cite{stead2005review}. The minute paper has students summarize concepts from a class period. Additionally, students could be asked to reflect on their remaining concerns or how their expectations have changed. This could be done anonymously at the end of class \cite{nilson2016teaching}. All of these approaches can be seen as a way to encourage students to think about what they are learning, how they are learning or as a means to provide formative feedback. Note that many of these techniques can easily be incorporated into any classroom environment and without an extensive effort on the part of the faculty member.

\subsection{Active Learning and its Use on Computer Science}
Studies involving active learning in the field of computer science date back at least 20 years with McConnell \cite{mcconnell1996active} reporting the use of active learning techniques across the breadth of the computer science curriculum. In particular, McConnell applied general techniques such as a modified lecture and think-pair-share as well as discipline specific techniques such as algorithm tracing. Chase and Okie also applied an active learning pedagogy to introductory computer science courses, although they referred to their approach as cooperative learning and peer instruction \cite{chase2000combining}. Through their study they found that the modified and peer based instruction reduced the percentage of students that withdrew from the course or received a 'D' or an 'F' and the approach seemed to have a stronger effect on the weaker students (i.e., there was not much change in grade between treatments for students in the 'B' or 'A' range) \cite{chase2000combining}. This reduction in failure rates mirrors what Freeman et al. found through their meta-analysis of active learning across STEM fields \cite{freeman2014active}. Additionally, interest and encouraged use by computer science faculty has been evidenced by community based online repositories such as Peer Instruction for Computer Science that contain information on research, professional development and prepackaged courses complete with active learning content (e.g., https://www.peerinstruction4cs.org/).

\subsection{Barriers to the Adoption of Active Learning Teaching Practices}
Evidence-based teaching practices (EBPs) are pedagogical approaches that have demonstrated value in the classroom \cite{trenshaw2017curmudgeon}. Active learning certainly falls into this category of teaching practices based on the meta-analysis of Freeman et al. In higher education, there are a number of reported barriers to the adoption of active learning and EBPs in general and these include institutional climate, time constraints and background \cite{michael2007faculty,turpen2016perceived}. It is interesting that many of these perceived barriers have been known for some time. In 2003, Michael held a two-day workshop on active learning to faculty. Among those present, reported barriers to active learning were broadly categorized as i) student characteristics (e.g., students do not know how to do active learning), ii) teacher characteristics (e.g., active learning requires too much preparation time) or iii) pedagogical issues (e.g., classrooms do not lend themselves to active learning) \cite{michael2007faculty}. In general, time and technology were the most cited significant barrier. Fast forward a decade and little has changed. After interviewing 35 physics faculty about their views regarding peer instruction, an evidence-based practice, and analyzing the results, Turpen et al. reported that time is still seen as a major constraint \cite{turpen2016perceived}.

To summarize, the literature is replete with studies that show the value of active learning in STEM fields in general and also in computer science. Active learning boosts students' classroom performance and experience. Nevertheless, adoption rates of active learning by faculty are arguably low and fall somewhere in the range of 20 to 40\% \cite{froyd2013estimates,grissom2017student}. Time and cost are often reported as barriers to the adoption of active learning \cite{michael2007faculty,park2014transformation,turpen2016perceived}. What seems to be lacking are the administrative and support staff's view as well as additional details into the background and attitudes of faculty. In particular, do administrators believe faculty already have the needed resources to implement active learning? If so, then it is not surprising that not much has changed with regards to active learning adoption. Are there other beliefs held by faculty that could be preventing them from making time to implement active learning? If so, then there could be additional barriers to be addressed. Given the prevalence and longevity of time as a barrier to adoption, are there activities that could be accomplished before faculty enter service that would lead to the use of active learning? If so, then perhaps a solution to the time barrier can be encountered.

\section{Methodology}

\subsection{Population}

The principle population for this study was computer science faculty and administration from public universities and colleges in a state located in the Upper Midwest. The email addresses of 369 computer science faculty were collected from public facing web portals. In addition to these faculty members, administrators responsible for instruction (e.g., dean of instruction, vice-provost for instruction, etc.) were also considered and sent an administrator survey. The survey was sent to 78 administrators and staff at community colleges, colleges and universities in the same state with email addresses harvested from public facing web portals.  

\subsection{Data Collection}
Data was collected via electronic surveys using SurveyMonkey. Access to both of the surveys expired on April 24, 2018.  An email invitation was sent out shortly after IRB approval was obtained and a reminder email was sent out two weeks after the initial email. The exact windows of data collection were February 26, 2018 to March 19, 2018 for the faculty survey and March 3, 2018 to March 25, 2018 for the administrator survey. The data was collected in an anonymous fashion and embedded functionally in SurveyMonkey was used to track who had submitted a response and who needed a reminder. Once the data collection had ended, data was exported from SurveyMonkey. The raw data was available in XLS format and then converted to a comma separated values (CSV) file. Scripts were created to convert the Likert scale descriptors to numerical values needed for analysis. From the faculty survey, 39 responses were received via the email invitations. Of these, 2 were largely incomplete and removed from consideration. This left 37 responses for a response rate of around 10\%. For the administrator survey, 16 responses were received, yielding a response rate of around 19\%. Overall, the response rate could be characterized as low and this limited the extent of the analyses that could be performed.  

\begin{table*}[h!]
  \caption{Demographic Information for Faculty Responses.}
    \label{tab:table1}
    \begin{tabular}{p{3cm} p{2cm} p{2cm} p{2cm} p{2cm} p{1cm}}
      Size of Institution & < 2000 & 2001 to 5000 & 5001 to 15,000 & 15,001 to 30,000 & > 30,000 \\
       & 2 & 1 & 14 & 6 & 13 \\
       \hline
      Type of Institution & Community College & \multicolumn{2}{l}{College/University (no or low research)} & \multicolumn{2}{l}{College/University (med. or high research)} \\
      & 0 & \multicolumn{2}{l}{4} & 32 \\
      \hline
      Position & Adjunct Faculty & \multicolumn{2}{l}{Instructor (fixed or renewable term)} & Tenure-Track & Tenured \\
      & 4 & \multicolumn{2}{l}{4} & 5 & 23 \\
      \hline
    \end{tabular}
\end{table*}

\begin{table*}[h!]
  \caption{Demographic Information for Administrator and Support Staff Responses.}
    \label{tab:table2}
    \begin{tabular}{p{3cm} p{2cm} p{2cm} p{2cm} p{2cm} p{1cm}}
      Size of Institution & < 2000 & 2001 to 5000 & 5001 to 15,000 & 15,001 to 30,000 & > 30,000 \\
       & 3 & 4 & 6 & 2 & 1 \\
       \hline
      Type of Institution & Community College & \multicolumn{2}{l}{College/University (no or low research)} & \multicolumn{2}{l}{College/University (med. or high research)} \\
      & 12 & \multicolumn{2}{l}{0} & 4 \\
      \hline
      Position & VP Instruction & Dean & Chair & \multicolumn{2}{l}{Director for Teaching Center} \\
      & 7 & 3 & 2 & 4 \\
      \hline
    \end{tabular}
\end{table*}

\subsection{Instrument}
There were two surveys that served as the primary data collection instruments.  One survey was sent to faculty and one was sent to administrators. The faculty survey contained four sections that included demographics, self-described teaching style, perceived barriers to change and mindset. The administrator survey was much shorter and only contained 3 sections that covered demographics, support services and barriers to teaching style change. The majority of the questions on the surveys were multiple choice, with most of the responses being on a Likert scale. Each survey had one opened end question at the end that asked respondents to state barriers to changes in teaching style.

Normal threats such as validity and reliability were in play in this study and were mitigated as best possible given the time constraints of the project. In this study, it was not possible to observe actual teaching styles. Given the increased discussion in the larger educator community regarding active learning, the questions regarding barriers to change were worded such that no mention of active learning was present. The questions to survey the teaching style used were worded in such a way that active learning, lecture and other loaded words were not used so as not to affect the participants' response (i.e., some words such as "lecture" have a negative connotation for many educators while others such as "active learning" may seem more modern or desired). These measures were taken to address some of the threats to validity.

\subsection{Rationale for the Questionnaires}
The survey instruments that were used in this study had not been used elsewhere (i.e., these were not repurposed from existing studies). There are advantages to using existing instruments, particularly if they have already been validated and reported to be reliable. Reusing an existing survey may also allow for an additional point of reference and comparison. In this case, it was not possible to find a suitable existing survey and custom surveys were constructed. 

Since the survey instruments used were specific to this study, their content warrants discussion. The first set of questions for the faculty survey largely collected demographic data. This included the size of the school, institution, type and position. Listed as responses were broad categories that were used to describe each of these based on personal experience. Also included were two questions that gauge interest and academic background in pedagogy. Collectively, these questions describe a faculty member's background.

The next set of questions on the faculty survey addressed attitudes for fixed versus growth mindset and locus of control. The questions over the locus of control were modified from Rose and Medway \cite{rose1981measurement}. The thought is that faculty who believe they have more control over student outcomes will be more interested in professional development or making changes to their teaching style. The rationale for the two questions over intelligence stemmed from a proposed study by Trenshaw \cite{trenshaw2017curmudgeon}. Here the thought is that faculty who believe that students can increase their intelligence are more likely to find and try techniques for those who might be struggling. Those who have more of a fixed mindset on intelligence may be less inclined to change their teaching style.

The third set of questions asked faculty to report on their teaching style. To limit the effect of existing connotations of lecture, peer instruction or active learning, the questions were worded such that they described classroom actions instead of a teaching style. Furthermore, the questions regarding lecture were worded such that they did not simply describe a Powerpoint presentation. These questions provided the basis for what the faculty member was currently doing in his or her classroom and were used to determine which faculty members were currently using active learning.

The fourth set of questions inquired about a faculty member's ability and desire to change his or her teaching style. It was thought that two questions (i.e., TS 2 and TS 4) may exhibit an inverse correlation since not sensing a need to change one's teaching style would seemingly indicated that the individual would not want to change his or her respective teaching style. Two questions asked about the faculty member’s perceived amount of support in terms of time and classroom environment. Clearly, this response depended on the specific changes made to the teaching style and the extent of interest if barriers to change were felt. Finally, faculty were asked about what resources they would believe helpful for changing their teaching style and what barriers might exist. 

The administrator survey was composed of three sets of questions. The first set of questions was demographical and mirrored the demographical information solicited from the faculty. This was done to possibly pair up administrators and faculty from similar institutions to further drill down at the level of institutional type or size.

The second set of questions for administrators was to gauge what barriers to the adoption of active learning may exist. These statements stemmed from barriers reported in the literature such as the time needed to make changes to a course or worries about the effects of active learning on student evaluations of teaching \cite{michael2007faculty}. Additional statements inquired about efforts to promote changes to teaching style such as communicating research to faculty or supporting faculty while they make changes. The statements were more direct and specific to active learning and evidence-based teaching practices than in the faculty survey.

The final set of questions for administrators mirrored the final set presented to faculty over perceived barriers. Some of the questions were arguably based on broad generalizations as administrators may not know if their faculty want to change their teaching style. Most likely some did and some did not but more importantly was the perception of the administrator to those statements since he or she makes decisions about supporting changes to teaching style from these very perceptions. The last question was an open-ended question about possible barriers faculty may encounter when changing their teaching style.

Overall, the intent was to keep the survey short. Faculty are already very busy and may be reluctant to report on their teaching habits. For this reason, an anonymous, electronic survey was used.

\section{Results}

The analysis applied to the data was descriptive statistics and plots of distributions. The ordinal values from the Likert scale were converted to an interval scale (i.e., 5 for strongly agree and 1 for strongly disagree) and then the mean, median and standard deviations were calculated for each response. Using R and the sjPlot package, the distribution of responses for each question was plotted \cite{sjplot,team2017r}. For data collected that was not on a Likert scale, the responses were categorized and counts were illustrated as word clouds. 

\subsection{Demographic Information about Respondents}

From the demographic data collected via the faculty survey, most of the respondents were tenured faculty from research oriented institutions. The distribution is shown in Table~\ref{tab:table1}. Additionally, 21 of the 36 respondents stated that they had attended a workshop or conference on teaching in the past 5 years and only 5 and 6 respondents stated that they had taken coursework related to teaching pedagogy as an undergraduate or graduate, respectively.

From the demographic data collected via the administrator and support staff survey, most of the respondents were vice-presidents of instruction or deans from community colleges. The distribution is shown in Table~\ref{tab:table2}. Additionally, 15 of the 16 respondents stated that they had attended a workshop or conference on teaching in the past 5 years and 3 and 10 respondents stated that they had taken coursework related to teaching pedagogy as an undergraduate or graduate, respectively.

\subsection{Quantitative Analysis and Results from Faculty Survey}
\begin{figure}
\includegraphics[height=3in, width=4in]{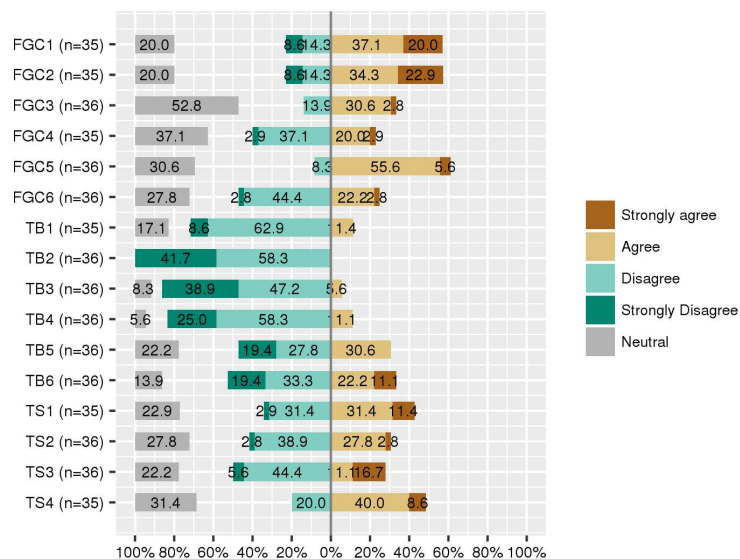}
\caption{Distribution of responses from faculty survey.}
\label{fig:faclik}
\end{figure}

The first analysis applied to the data was descriptive statistics. The ordinal values from the Likert scale were converted to an interval scale (i.e., 5 for strongly agree and 1 for strongly disagree). Then the mean, median and standard deviations were calculated for each response. Tables~\ref{tab:table3}, ~\ref{tab:table4},~\ref{tab:table5} summarize descriptive statistics for each question on the survey. Using R and the sjPlot package, the distribution of responses for each question was plotted and these plots constitute Figure~\ref{fig:faclik} \cite{sjplot,team2017r}.

\begin{table}[h!]
  \caption{Mean, Standard Deviation and Median for Questions for Computer Science Faculty Regarding Student Attitudes and Aptitudes}
  \label{tab:table3}
  \begin{tabular}{p{4cm} c c c}
  \textbf{Question} & \textbf{Mean} & \textbf{Std. Dev.} & \textbf{Median} \\
  \hline
  FGC1) While you can learn new things you cannot change
your basic intelligence. & 2.54 & 1.22 & 2 \\
  \hline
  FGC2) Your intelligence is something that cannot be changed
much. & 2.51 & 1.25 & 2 \\
  \hline
  FGC3) If the students in your class perform better than they
usually do on an exam, then this is because you did a better
job of teaching the content area rather than because the
students worked harder. & 2.78 & 0.72 & 3 \\
  \hline
 FGC4) If a student fails an exam, it is more likely because the
student did not attend class rather than that you did not
provide an adequate number of examples. & 3.17 & 0.89 & 3 \\
  \hline
  FGC5) If one of your students could not complete a
homework assignment it is more likely that you gave an
assignment that was too difficult rather than that the student
was not paying attention. & 2.42 & 0.73 & 2 \\
  \hline 
  FGC6) Suppose you are teaching a programming construct to
a student who has trouble learning it, this is most likely
because you could not explain it very well rather than the
student was not able to understand it. & 3.22 & 0.93 & 3 \\
 \hline
  
  \end{tabular}
\end{table}

\begin{table}[h!]
  \caption{Mean, Standard Deviation and Median for Questions for Computer Science Faculty Regarding Teaching Techniques}
  \label{tab:table4}
  \begin{tabular}{p{4cm} c c c}
  \textbf{Question} & \textbf{Mean} & \textbf{Std. Dev.} & \textbf{Median} \\
  \hline
   TB1) I guide my students through course topics as they take
notes and follow along. & 3.69 & 0.8 & 4 \\
  \hline
   TB2) I provide many examples and aids via slides or other
in-class presentations.& 4.42 & 0.5 & 4 \\
  \hline
  TB3) I frequently ask students to respond to in-class
questions or prompts. & 4.19 & 0.82 & 4 \\
  \hline
  TB4) I routinely solicit feedback, written or verbal, from
students during class.& 3.97 & 0.88 & 4 \\
  \hline
   TB5) I structure class so that students talk to one another
about course concepts. & 3.36 & 1.13 & 3 \\
  \hline 
  TB6) I require that students regularly work together in small
groups. & 3.28 & 1.33 & 4 \\
 \hline
  
  \end{tabular}
\end{table}

\begin{table}[h!]
  \caption{Mean, Standard Deviation and Median for Questions for Computer Science Faculty Regarding Changes to Teaching Style}
  \label{tab:table5}
  \begin{tabular}{p{4cm} c c c}
  \textbf{Question} & \textbf{Mean} & \textbf{Std. Dev.} & \textbf{Median} \\
  \hline
  TS1) I have sufficient time to make desired changes to my
teaching style. & 2.83 & 1.1 & 3 \\
  \hline
  TS2) I would like to change my teaching style. & 3.11 & 0.95 & 3 \\
  \hline
  TS3) My existing classrooms support any changes I would
like to make to my teaching style. & 3.11 & 1.21 & 3.5 \\
  \hline
  TS4) I see no reason to make changes to my teaching style. & 2.63 & 0.91 & 3 \\
  \hline
  
  \end{tabular}
\end{table}

There are a few interesting patterns that are evident from the distribution of responses. First, questions TB1 through TB6 were proxy questions to ascertain what faculty were currently doing in the classroom. Questions TB1 and TB2 describe a more traditional or passive teaching style and the remaining describe a more active teaching style in line with evidence-based teaching practices (e.g., providing formative feedback in class, peer instruction, etc.). Interestingly, the respondents overwhelming agreed with statements TB1 through TB4 meaning that they did see their teaching style as traditional and passive and that they also provided formative feedback in class. There is also roughly 50\% of the respondents who reported that they did not encourage students to work in groups or talk about course concepts in class. Thus, there was a seemingly large portion of the respondents who did not characterize their classroom as active. To see if there were any differences in the distribution of responses based on previous coursework, plots were generated from the responses of only those who had reported previous coursework in pedagogy (not shown here). While the numbers are limited, almost all of these respondents reported making use of evidence-based practices such as peer instruction and in-class formative feedback. These individuals also placed more control and responsibility for learning on the part of the instructor or took a neutral position, as indicated from their responses on FGC1 through FGC6.

Next, in terms of locus of control, the vast majority of the respondents (i.e., FGC3: 86.2\%, FGC4: 77.1\%, FGC5: 91.8\%) indicated that were neutral or leaned towards attitudes that placed the onus on the student to improve classroom performance (or were at least reluctant to place responsibility on themselves). This is important as faculty will be less likely to change their teaching style if they believe what they do has little effect. The respondents also largely agreed with the idea that students have a certain level of raw talent (i.e., intelligence) that is fixed but are capable of learning new things.

In terms of changing teaching style, it was refreshing to see that at least 30\% of the respondents felt as though they had desire and time to change their teaching style. A large portion, however, did not believe that their existing classrooms would support the changes that they would like to make. More than 50\% of the respondents did not feel the need to change their teaching style and this is not necessarily a negative since they could have already be making use of evidence-based teaching practices and active learning.  In doing so, it was revealed that there were many who reported a desire to stay with passive delivery. Perhaps more favorably, of those who disagreed with statement of not wanting to change their teaching style (i.e., those who seemingly reported a desire to change their teaching style), many also reported making use of passive pedagogy. This is important as they would seemingly be turning away from passive styles when they make changes to their teaching style. Still, the numbers are too small to make any strong generalizations.

The last closed form question presented to the participants was a list of aids that faculty might see as helpful for changes in teaching style. The list included a workshop, colleague, peer learning cohort, and an institutional resource such as a center of teaching. More than one selection was allowed. Table~\ref{tab:table6} summarizes the counts by resource type and also provides a break down by type of position. A workshop seems to be preferred support with a cohort being the least preferred. Selections may correlate with the perceived amount of effort required for each resource but that point would need to be further investigated.

\begin{table}[h!]
 \caption{Helpful Resources for Those Wanting to Change their Teaching Style}
 \label{tab:table6}
 \begin{tabular}{p{2cm} p{1cm} p{1cm} p{1.5cm} p{1.5cm}}
  & Workshop & Colleague & Peer learning cohort & Institutional resource \\
  \hline
  Tenured (n=23) & 15 & 14 & 5 & 14 \\
  \hline
  Tenure-track (n=5) & 5 & 2 & 2 & 2 \\
  \hline
  Instructor (n=4) & 1 & 1 & 1 & 2 \\ 
  \hline
  Adjunct (n=4) & 3 & 2 & 2 & 2 \\
  \hline
  Total (n=36) & 24 & 19 & 10 & 20 \\
  \hline
  \end{tabular}
\end{table}

\subsection{Qualitative Analysis and Results from Faculty Survey}

\begin{figure}

\includegraphics[height=2in, width=3in]{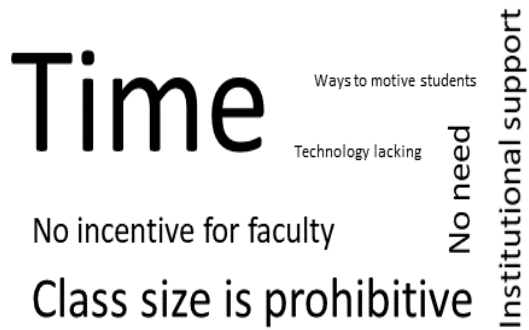}
\caption{A word cloud of commonly occurring words for barriers to changing teaching style.}
\label{FacultyCloud}
\end{figure}

A simple qualitative analysis was also completed on an open-ended question at the end of the survey which asked "What barriers, if any, exist to changing your teaching style?" (e.g., TS6). This analysis looked at common themes that appeared in the written responses. The most common theme was time, with faculty expressing concerns about the time needed to change their teaching style. The number of students or class size was the second most commonly provided response. A word cloud counting the occurrence of themes in written responses and then adjusting the size of the word in a manner proportional to its appearance. The word cloud was made manually and included as Figure~\ref{FacultyCloud}.

\subsection{Quantitative Analysis and Results from Administrator Survey}

\begin{table}[h!]
  \caption{Mean, Standard Deviation and Median for Questions for Administrators and Support Staff Regarding Available Teaching Support}
  \label{tab:table7}
  \begin{tabular}{p{4cm} c c c}
  \textbf{Question} & \textbf{Mean} & \textbf{Std. Dev.} & \textbf{Median} \\
  \hline
AS1) Your faculty have been made aware that the
administration understands the effects of evidenced based
teaching practices (e.g., active learning teaching styles) on
student evaluations of teaching for the purposes of tenure or
reappointment.  & 3.5 & 0.89 & 4 \\
  \hline
 AS2) Your faculty have access to on-campus resources to
faculty who desire to change their teaching style.  & 4.69 & 0.48 & 5 \\
  \hline
 AS3) Your on-campus facilities can support faculty who wish
to change their teaching styles.  & 4.44 & 0.51 & 4 \\
  \hline
  AS4) Your faculty have opportunities to interact with other
faculty on-campus to support development or implementation
changes to teaching styles. & 4.63 & 0.5 & 5 \\
  \hline
  
  \end{tabular}
\end{table}

\begin{table}[h!]
  \caption{Mean, Standard Deviation and Median for Questions for Administrators and Support Staff Regarding Changes to Teaching Style}
  \label{tab:table8}
  \begin{tabular}{p{4cm} c c c}
  \textbf{Question} & \textbf{Mean} & \textbf{Std. Dev.} & \textbf{Median} \\
  \hline
 TS1) Your faculty have sufficient time to make desired
changes to their teaching style. & 3.19 & 1.67 & 3 \\
  \hline
  TS2) Your faculty would like to make changes to their
teaching style. & 3.25 & 0.58 & 3 \\
  \hline
  TS3) My institution’s existing classrooms support changes
my faculty would like to make to their teaching style. & 3.56 & 0.81 & 4 \\
  \hline
  TS4) I see no reason why my faculty should make changes
to their teaching style. & 1.88 & 0.72 & 2 \\
  \hline
  
  \end{tabular}
\end{table}

\begin{figure}

\includegraphics[height=3in, width=4in]{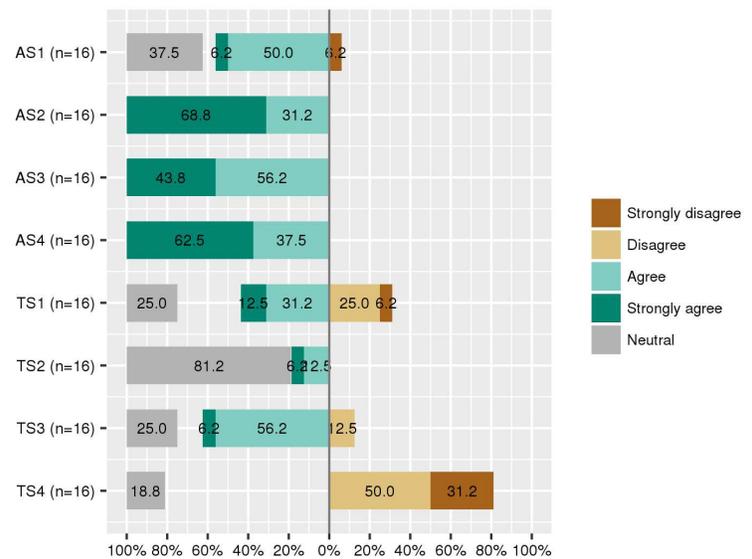}
\caption{Distribution of responses from administrator and support staff survey.}
\label{AdminLikerts}
\end{figure}

The quantitative analysis performed on the administrator and support staff survey was similar to that of the faculty survey. The first analysis applied to the data was descriptive statistics. The ordinal values from the Likert scale were converted to an interval scale (i.e., 5 for strongly agree and 1 for strongly disagree). Then the mean, median and variance were calculated for each response. Tables ~\ref{tab:table7} and~\ref{tab:table8} summarize descriptive statistics for each question on the survey. Using R and the sjPlot package, the distribution of responses for each question was plotted and these plots constitute Figure~\ref{AdminLikerts} \cite{team2017r,sjplot}.

The data, while limited, has clear indications from the respondents. First, administrators and support staff believed that their faculty had the resources they needed to change their teaching style. All respondents either agreed or strongly agreed with statements A2, A3 and A4. Respondents also largely agreed that their faculty had been made aware of evidence-based teaching practices. Administrators and support staff largely disagreed with the statement TS4, "I see no reason why my faculty should make changes to their teaching style". They also responded that their facilities would support change. The only statement in which there was some disagreement was if faculty had time to make desired changes. Thus, administrators and support staff were indicating that faculty are aware of evidenced based teaching practices and have what they need to make changes to their teaching style. Table~\ref{tab:table9} tallies the resources that administrators believed would be helpful resources to faculty that desire to change their teaching style. There the distribution of the helpful resources was more uniform than what was reported by faculty.

\begin{table}[h!]
 \caption{Helpful Resources for Those Wanting to Change their Teaching Style}
 \label{tab:table9}
 \begin{tabular}{p{2cm} p{1cm} p{1cm} p{1.5cm} p{1.5cm}}
  & Workshop & Colleague & Peer learning cohort & Institutional resource \\
  \hline
  Chair (n=2) & 2 & 2 & 1 & 2 \\
  \hline
  Dean (n=3) & 2 & 3 & 2 & 1 \\
  \hline
  VP Instruction (n=7) & 6 & 5 & 7 & 6 \\ 
  \hline
  Dir. Teaching Center (n=4) & 3 & 3 & 4 & 3 \\
  \hline
  Total (n=16) & 13 & 13 & 14 & 12 \\
  \hline

  \end{tabular}
\end{table}

\subsection{Qualitative Analysis and Results from Administrator Survey}

\begin{figure}

\includegraphics[height=2in, width=3in]{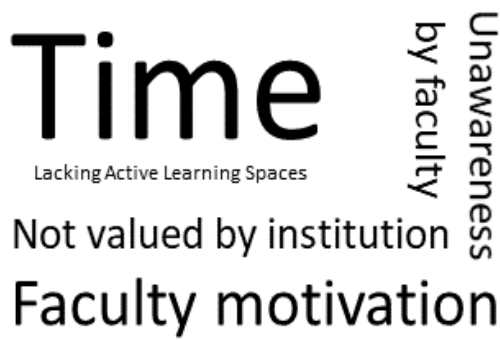}
\caption{A word cloud of commonly occurring words used by administrators for describing barriers their faculty face to changing teaching style.}
\label{AdminCloud}
\end{figure}

A qualitative analysis similar to that performed on the faculty survey was completed for an open-ended question at the end of the administrator and support staff survey. The question asked "What barriers, if any, do you believe exist for your faculty that desire to change their teaching style?" (e.g., TS6). This analysis looked at common themes that appeared in  the written responses. The most common theme again was time. Faculty motivation was the second most commonly provided response. A word cloud counting the occurrence of themes in written responses and then adjusting the size of the word in a manner proportional to its appearance. The word cloud was made manually and included as Figure~\ref{AdminCloud}.

\section{Discussion}

This work was motivated in part by two questions: i) what is the relationship between usage of active learning and faculty's attitudes, beliefs and background? and ii) what is the relationship between the barriers to change in teaching style as perceived by computer science faculty and administration? Through an electronic survey sent to computer science faculty and administrators and support staff, some light has been shed on both questions.

In terms of attitudes, background and beliefs, there was a clear pattern that those with prior exposure to coursework in pedagogy also make use of active learning techniques and place onus on the educator to improve student outcomes. The data is very limited and there is no implication of causality but this relationship needs to be explored more. When looking generally at all of the respondents, faculty more readily attributed academic failures to students rather than actions taken by the educator. This is important because if an educator believes educational outcomes rest largely with the student, he or she may be less likely to try other teaching styles that might be more effective. Another interesting pattern was between those reported using passive learning techniques and changes to teaching style. In particular, of those who disagreed with statement of not wanting to change their teaching style (i.e., those who seemingly reported a desire to change their teaching style), many also reported making use of passive pedagogy. This may indicate that they did not see the value in their current teaching style.

The relationship between faculty and administrative perceptions of barriers to changes in teaching style is multi-faceted. There was one common, clear point of agreement that was not revelatory in nature and this was time. Both faculty and administrators agreed that time was the largest barrier for faculty that want to change their teaching style. Faculty and administrators had differing views of what is needed to support faculty desiring to make changes to teaching style with administrators and support staff indicating that faculty had everything they needed with the possible exception of time. From the administrators' standpoint, faculty have everything they need to change and have been made aware reasons to change. If they do not, then it must be a question of motivation. Faculty, on the other hand, listed some reservations about the classrooms and if they could accommodate their changes. Faculty also expressed some clear preferences on how to support faculty that wanted to change their teaching style (i.e., workshops) while administrators took a more broader approach and indicated that most of the suggested resources would be helpful.

\subsection{Recommendations for Future Research}
In terms of directions for future research, two recommendations are i) to investigate new avenues to provide future faculty with exposure to active learning pedagogy in their undergraduate and graduate training when they may have more time and flexibility and ii) solicit responses from the broader community of computer science educators. From the faculty survey, it is clear that time continues to be a barrier for faculty that desire to make changes their teaching style. This manifested itself as the most common theme from the open-ended responses and also by means of the most commonly suggested support, namely workshops. Workshops require less time and commitment than a peer cohort. Administrators and support staff also recognized time as a barrier but interestingly administrators felt that the even more intensive support systems such as peer cohorts would be useful to faculty.

Intuitively, time, or lack of it, as a barrier to change makes sense regardless of the type of institution. Faculty at smaller, teaching oriented institutions likely have a higher teaching loads and faculty are more research oriented institutions may be directed to focus their efforts on research. While only mentioned by a few faculty members and administrators via the free response questions, it was clear that there often is little incentive provided by an institution to change teaching pedagogy or motivate faculty to improve teaching. 

In light of the limitations of time and incentive to change teaching style that face faculty in the field, the best option may be to reach future faculty. Few of the faculty stated that they had taken any formal course work in teaching pedagogy as undergraduate or graduate students. While graduate students are busy as well, they arguably have more time and flexibility than faculty. Graduate minors or certificates exist in college teaching but these may require a greater commitment than future faculty can make. 

A more general recommendation for this area of research is the need to solicit responses from a broader range of educators. For survey based studies, and particularly for electronic or remote (i.e., not face-to-face) surveys, selection bias will be an issue. In this study, a question was included that asked about prior attendance to a workshop on teaching or education. Of the respondents in this study, the percent who answered affirmatively was over 50\% and this seems high for computer science faculty. Selection bias was listed as a limitation in many of the studies reviewed but few included a proxy to gauge it. Additional studies that are designed to access this differing view are needed. Is it really laziness, as one administrator stated, that keep faculty form changing their teaching style or are there questions about the efficacy of change on the part of these faculty? Do these faculty think students are beyond help? These are important questions that need to be addressed.

\subsection{Limitations and Cautions}

Some caution should be taken when attempting to generalize these results. First, from the demographic information collected it appears that the responses collected were skewed towards tenured faculty members and faculty from medium to large research institutions. Faculty at this rank and at research oriented institutions may feel less pressure to explore teaching pedagogy or teaching style. Second, almost two-thirds of the respondents reported that they had attended a conference or workshop on teaching in the past 5 years. From anecdotal experience, this percentage seems high and since this question was used a proxy gauge existing interest in pedagogy, it could indicate the results suffer from self-selection bias. In particular, the results could be biased towards those already predisposed an interest in pedagogy and to the value of evidence-based teaching practices.  Finally, the number of respondents was low (e.g., ~10\% of faculty invited). 

\section{Conclusion}

At the root of this study was an investigation to identify barriers to active learning. While the value of active learning was known, faculty adoption has remained low. Faculty and administration are in agreement that time is a major barrier for faculty that desire to change teaching style. This is disappointing given that faculty have reported their concerns over time limitations for over 15 years and it remains a barrier today. This is indicative that a different approach is needed to increase student access to active learning and evidence-based teaching practices as faculty simply do not have time to adapt when in service. Providing pedagogical training during graduate school could address the barrier of time and this study provides some preliminary evidence that pedagogical training completed as a student may correlate to active learning use in the classroom. Of those that did express a desire to change their teaching style, many were making use of passive learning in their classrooms and there were indications that many faculty want to change their teaching style.

\bibliographystyle{ACM-Reference-Format}
\balance
\bibliography{acmart.bib} 

\end{document}